\documentclass[man]{apa7}
\usepackage[american]{babel}
\usepackage{lineno}
\usepackage{csquotes}
\usepackage[style=apa,sortcites=true,sorting=nyt,backend=biber]{biblatex}
\DeclareLanguageMapping{american}{american-apa}
\title{Redrawing attendance boundaries to promote racial and ethnic diversity in elementary schools}
\shorttitle{Redrawing school attendance boundaries}

\author{Nabeel Gillani,$^{1\ast}$ Doug Beeferman,$^{2}$ Christine Vega-Pourheydarian,$^{2}$ Cassandra Overney,$^{2}$ Pascal Van Hentenryck,$^{3}$ Deb Roy$^{2}$}
\affiliation{$^{1}$Northeastern University $^{2}$Massachusetts Institute of Technology $^{3}$Georgia Institute of Technology}

\authornote{To whom correspondence should be addressed:  n.gillani@northeastern.edu.}

\abstract{

Most US school districts draw "attendance boundaries" to define catchment areas that assign students to schools near their homes, often recapitulating neighborhood demographic segregation in schools.  Focusing on elementary schools, we ask: how much might we reduce school segregation by redrawing attendance boundaries?  Combining parent preference data with methods from combinatorial optimization, we simulate alternative boundaries for 98 US school districts serving over 3 million elementary-aged students, minimizing White/non-White segregation while mitigating changes to travel times and school sizes.  Across districts, we observe a median 14\% relative decrease in segregation, which we estimate would require approximately 20\% of students to switch schools and, surprisingly, a slight \textit{reduction} in travel times.  We release a public dashboard depicting these alternative boundaries (\url{https://www.schooldiversity.org/}) and invite both school boards and their constituents to evaluate their viability.  Our results show the possibility of greater integration without significant disruptions for families.

}

\begin{document}
\maketitle

\section{Introduction}

It has been over 65 years since the US Supreme Court ordered the racial desegregation of schools~\parencite{brownvboard}.  Yet segregation by race and income in K12 schools continues to hamper access to quality education for millions of children across the US~\parencite{reardon2018testgaps}, despite strong evidence that integration reduces achievement gaps between lower income students of color and their more affluent, majority race counterparts~\parencite{billings2013cms, johnson2011desegregation, wells1994integration}.  Of course, increasing diversity by fostering more demographic integration is not a foolproof method for reducing achievement gaps.  Too often, even after addressing segregation at the school level, segregation persists at the classroom or friendship level~\parencite{potter2016stamford, card2016gt, tatum1997cafeteria, moody2001tracking}, or low income students of color feel unsupported in more integrated environments~\parencite{comer1988integration}.  Diversity done wrong can cause more harm than good.  And yet, more diverse schools can serve as a necessary first step toward providing children from different racial and socioeconomic backgrounds the chance to mix and learn from one another.  This learning and mixing is important beyond its potential role in reducing achievement gaps: it can also help increase empathy, compassion, reflective thought~\parencite{wells2016benefit}, and encourage more welcoming attitudes towards diversity later on in life~\parencite{wells1994integration, davies2011intergroup}.  There is evidence to suggest that all students can benefit from racially and socioeconomically diverse classrooms.

Yet across the US, the vast majority of students attend the schools closest to their homes by virtue of how ``school attendance boundaries''---or catchment areas---are drawn~\parencite{monarrez2021boundaries, richards2014gerrymandering, saporito2016shapes}, leading schools to recapitulate neighborhood-level segregation by race and income.  Despite the impact attendance boundaries can have on racial and ethnic diversity in schools, most school segregation results from how the lines \textit{between} districts are drawn (e.g. separating cities from suburbs), instead of school-specific boundaries \textit{within} districts~\parencite{fiel2013boundaries, monarrez2021boundaries}.  Redrawing district boundaries is arguably a more difficult problem, however, because it falls under the purview of state legislatures—making it subject to the whims, frictions, and bureaucratic inefficiencies of similarly-contentious political issues manifesting at state and federal levels.  On the other hand, changing attendance boundaries within districts generally falls under the purview of those districts.  Indeed, a landmark 2007 Supreme Court case outlawed the use of individual students’ racial backgrounds as an input into school desegregation efforts and effectively encouraged districts to explore the redrawing of school attendance boundaries as a desegregation policy~\parencite{totenberg2007supreme}.

The purpose of this paper is to explore to what extent it might be possible to do this---i.e., redraw attendance boundaries within districts in order to achieve more diverse schools---without imposing large travel burdens, overcrowding schools, or fragmenting existing geographic ``cohesion'' (i.e., contiguity).  We frame our inquiry as an constrained optimization problem and ask two overarching questions: 1) how can we re-assign geographies to schools in order to minimize racial segregation while respecting parents' travel time and class size preferences?  And 2) how fairly are these reductions in segregation, and associated costs---namely, changes in travel times and school switching requirements---distributed across Asian, Black, Hispanic/Latinx, Native American, and White students?  To explore these questions, we focus on elementary schools for similar reasons as~\parencite{monarrez2021boundaries}: because their boundaries often approximately combine to form the boundaries of the middle and high schools they ``feed'' to, and hence, are foundational in shaping diverse exposures at an early age.  We use parent input and computational tools to simulate changes across 98 large school districts across the US with district elementary schools that are classified as non open-enrollment: that is, attendance at these schools are entirely a function of which neighborhoods are zoned to attend them.  These schools collectively serve over 3 million students.  

Our findings show that alternative attendance boundaries could produce a relative decrease of 12\% in White/non-White segregation across districts.  These boundaries would require nearly 20\% of students to switch schools, and interestingly, a slight \textit{decrease} of just under one minute in these students' time spent traveling to school. On average, these ``costs'' of added diversity appear to be fairly distributed across different student groups, though through two case studies, we see that this can vary by district and rezoning.  We release our code, several datasets, and a public dashboard (\url{https://www.schooldiversity.org/}) inviting interested researchers and school districts across the US to further explore the opportunities and potential trade-offs involved in changing attendance boundaries to advance integration objectives.  In the subsection below, we offer additional background on the topics of boundary-based school assignment and attendance boundary changes. 

\subsection{Background on attendance boundary-based school assignment}
The expansion of school choice programs has sought to challenge the geographic determinism of boundary-driven school assignment and thereby also mitigate school segregation~\parencite{kahlenberg2016integration}.  However, choice too, has been shown in several instances to perpetuate segregation due to self-selection of certain families into certain schools~\parencite{monarrez2022charters,whitehurt2017segregation,candipan2019neighborhoods}.  Boundaries continue to play a prominent role in student assignment: as of 2016, approximately 20\% of students in grades 1-12 participated in some type of public school choice (including 8\% opting for charter schools); 9\% attended private schools; and the remaining 71\% attended an assigned school, likely determined by geography~\parencite{nces2021choicefacts}.  Choice programs have continued to gain popularity in recent years, particularly as some subsets of families have sought new avenues for mitigating the pandemic's effects on their children's learning~\parencite{houlgrave2021choice}, yet place-based school assignment continues to be the norm.  Even in choice settings, where students live might influence the priority they are assigned to attend a certain schools~\parencite{monarrez2021urban}, or even which schools are part of the choice set~\parencite{campos2022zones}.  This makes attendance boundaries, and more generally, place of residence a perennially important factor in school attendance policies.  The implications of these boundaries and resultant segregation can run deep: for example, they have been shown to demarcate stark gradients in access to gifted and talented programs, quality teachers, school counselors, and a number of other educational resources~\parencite{monarrez2021urban}.  

Still, changing attendance boundaries within districts continues to be a highly contentious topic, especially when issues of diversity are also at stake~\parencite{mcmillan2018boundaries}.  Parents may fear that rezoning students will increase travel times through longer ``busing''~\parencite{frankenberg2011polls}, reduce quality of education~\parencite{zhang2008flight}---which they often define vis-á-vis test scores~\parencite{abdulkadiroglu2019parents} and class sizes~\parencite{gilraine2018class}, produce unsafe school environments~\parencite{baltimore2019}, drop property values~\parencite{kane2005housing, bridges2016eden, black1999housing}, fragment communities~\parencite{bridges2016eden, baltimore2019}, and require a number of other sacrifices.  

These concerns, while sometimes reasonable, often impede practical paths towards achieving more diverse and integrated schools---e.g., by sparking ``white flight'' in response to unfavorable school assignment policies~\parencite{reber2005flight} and souring public opinion towards desegregation efforts as a result of concerns about long-distance busing and other inconveniences~\parencite{delmont2016busing}.  Furthermore, despite parents increasingly expressing support for school integration through polls and surveys~\parencite{frankenberg2011polls,torres2020integration}, they continue to ``vote with their feet'', deciding where to live and send their children to school in ways that reflect racialized preferences~\parencite{billingham2016parents,hailey2021parents,hall2017migration,charles2003seg,iceland2010households}.  Such preferences, especially when aggregated and compounded across families, can yield extreme levels of segregation across neighborhoods, cities, and schools~\parencite{card2008tipping,schelling1971segregation}.  Shifting these underlying preferences is one of the greatest challenges of our time, and is critical for the implementation of sustainable school desegregation efforts that persist in the face of changing legal mandates~\parencite{billings2013cms}.  Alongside this deeper work, however, it is also critical to identify if there are pathways to achieving more diverse and integrated schools \textit{today}---in the case of our focus, through alternative attendance boundaries---that families may earnestly consider and not immediately dismiss because they significantly disrupt and decrease day-to-day quality of life.

This is empirical question is what motivates our current study, which is a simulation-based exploration of how much alternative attendance boundaries within districts might reduce racial and ethnic segregation, subject to various constraints.  While several studies have explored relationships between attendance boundaries and school segregation~\parencite{monarrez2021boundaries, saporito2016shapes, richards2014gerrymandering}, we have found few that have explored actually changing school boundaries---with the exception of~\parencite{mota2021districts,caro2004integer,liggett1973optimization,clarke1968optimization}---yet these have not focused on achieving greater racial and ethnic diversity across schools as the main objective of their approach.  Larger districts may hire external vendors to explore alternative boundary scenarios; however, their exact tools and methods are often opaque, and diversity is rarely, if ever, a primary objective---though it is sometimes included as a constraint or post hoc measure~\parencite{mcps2021analysis}.  To our knowledge, our work is the first to simulate alternative attendance boundaries optimized to achieve racial and ethnic desegregation across a large number of US school districts.  Simulations alone are not sufficient to drive policy change, especially in the face of parents and others who might oppose such change, but may help illuminate possible paths to integration ``within reach'' that both districts and families may not have previously explored.

We focus on White/non-White segregation as our primary quantity of interest given its historical significance within the US and abroad; its association with other family-level factors that have been shown to correlate with educational outcomes, like socioeconomic status~\parencite{reardon2018testgaps}; and the precision and reliability with which racial/ethnic data is available at the granularity of schools and small geographic units like Census blocks (as opposed to measures of socioeconomic status among parents, which are also critical in the discussion about school segregation, but less reliably and precisely defined and available~\parencite{harwell2010frl}).  White/non-White segregation does not perfectly capture patterns of segregation across all school districts: for example, in some district settings, White and Asian students may be more likely to attend schools together, segregated away from their Black and Hispanic/Latinx counterparts~\parencite{chang2018shsat}.  Nevertheless, across most districts, including those in our sample, White, Black, and Hispanic/Latinx students constitute the vast majority of the population, rendering White/non-White segregation an important dimension of analysis.     

\section{Data and Methods}

\subsection{Optimization model}

We explore the extent to which we might reduce segregation across three different metrics: the widely-used Dissimilarity index ($D$); the related Gini index ($G$); and the Variance Ratio index ($V$).  All three metrics are presented and discussed in~\parencite{massey1988segregation}, with formal definitions below:

\begin{equation}
        \label{eq:1}
        D = \frac{1}{2}~\Sigma_{s \in S} ~|\frac{w_s}{w_T} - \frac{t_s - w_s}{T - w_T}|
\end{equation}

\begin{equation}
        \label{eq:2}
        G = \frac{\Sigma_{s1 \in S}~\Sigma_{s2 \in S} ~({t_{s1} t_{s2}})~|\frac{w_{s1}}{t_{s1}} - \frac{w_{s2}}{t_{s2}}|}{2 T^2 \left[\frac{w_T}{T}\right]\left[1 - (\frac{w_T}{T})\right]}
\end{equation}

\begin{equation}
        \label{eq:3}
        V = \frac{\left[~\Sigma_{s \in S}~(\frac{w_s}{w_T})(\frac{w_s}{t_s})\right] - \left[\frac{w_T}{T}\right]}{1 - (\frac{w_T}{T})}
\end{equation}

Here, $s$ is an elementary school across all district elementary schools $S$; $t_s$ and $w_s$ indicate the total and total White students at $s$, respectively; and $T$ and $w_T$ indicate the total and total White students across the district, respectively.  Perfectly integrated districts---where the proportion of White/non-White students in each school reflects district-wide proportions---would receive a score of 0 under these measures, while perfectly segregated districts would receive a score of 1.  

The purpose of exploring several different measures of segregation is that each describes something slightly different about how students from different racial and ethnic backgrounds are distributed across schools.  Furthermore, each has its own merits and pitfalls.  Dissimilarity has historically been the most widely used measure of segregation and represents the proportion of White students in the district who would need to switch schools in order to achieve perfect integration~\parencite{jakubs1977dissim}.  Yet it also suffers from a number of shortcomings, namely, its 1) failure to fully respect the ``transfers/exchanges'' principle, whereby movement of students from schools with a higher proportion of other same-race students to a school with a lower proportion may not decrease dissimilarity unless one school is over-represented, and the other under-represented, with respect to the group's district-wide prevalence~\parencite{james1985seg}; and 2) potential equal treatment of changes that lower the index, even if some may have more normative value than others (like reducing a school's demographic population of 100\% to 90\% belonging to a certain group, vs. 60\% to 50\%)~\parencite{winship1978dissim}.  The Gini index is closely related to the dissimilarity measure but respects the transfers/exchanges principle.  The variance ratio index (also known as the normalized exposure index in multi-group settings~\cite{owens2022normalized}) is essentially an isolation index that accounts for the underlying demographic distribution of the given district.  In our context, it indicates how much higher the fraction of White students is in the average White student's school compared to the average non-White student's school.  The variance ratio index also respects the transfers/exchanges principle, but does not respect the principle of ``composition invariance'': e.g., doubling the number of non-White students in each school would decrease the variance ratio index (because it would increase White/non-White exposures), even though it may not necessarily change the extent to which school-level proportions differ from district-wide proportions (which $D$ and $G$ more closely measure).  Nevertheless, it is a popular measure of segregation in part because it offers insight into what the average-case encounters between students from different backgrounds might be.  

There are many other valid measures of segregation, including multi-group measures like Theil's Entropy Index~\parencite{reardon2002multigroup}, and we invite interested readers to build upon our code (which we release with this paper) to explore these and other measures further.  Critically, we note that all of these measures of segregation are naive in that they do not account for within-school segregation and sorting~\parencite{moody2001tracking, tatum1997cafeteria}---including levels of ``friending bias''~\parencite{chetty2022socialcapitalII} that may manifest within schools and subsequently affect who connects with whom, how social capital is shared, and ultimately, the extent to which more diverse schools translate into more engagement across lines of difference.

With these considerations in hand, we design a rezoning algorithm that seeks to re-assign Census blocks to elementary schools within each district in order to minimize each of the above measures.  Rezoning problems are generally computationally challenging because of the many geographic units they operate over, and the sometimes large number of constraints (e.g., in the case of contiguity constraints) they impose.  Much redistricting work to date has focused on congressional redistricting, and many approaches to this have used mixed-integer programming (MIP) as a core building block~\parencite{becker2020redistricting}, often augmented with problem-specific search strategies~\parencite{gurnee2021fairmandering}.  To compute these combinatorial optimization problems---which are ``NP-hard'' and lack efficient, polynomial-time solutions---we use constraint programming~\parencite{pascal1989cp} via the CP-SAT model in Google's Operations Research (OR) Tools library~\parencite{cpsat}, which has been shown to perform extremely well on a number of different types of combinatorial optimization problems~\parencite{cpsat2020youtube}.  Constraint programming enables us to more flexibly express constraints and nonlinear objective functions that may otherwise be difficult to encode.  While CP-SAT is able to find high-quality solutions to these notoriously difficult geographic rezoning problems, given the size of most districts, it is generally unable to prove that the discovered solutions are optimal.  This means that it may be possible to improve upon the reductions in segregation we report, perhaps through additional computational resources and/or alternative model and solver specifications. 

The algorithm factors in the following constraints, given they represent topics that are often top of mind for parents and district officials when exploring boundary changes~\parencite{mcmillan2018boundaries,mcps2021analysis}:  

\begin{enumerate}
    \item \textbf{Maximum travel time increases}.  We use the OpenRouteService API~\parencite{ors2022} to estimate driving times from Census blocks to schools (see more below), and require that re-assignments of blocks to new schools do not increase estimated travel times by more than X\% for any given family.
    \item \textbf{Maximum school size increases}.  We use the total population at a given school as a proxy for a quantity parents often care about in their children's schools---class sizes~\parencite{gilraine2018class}---and require that this total does not exceed Y\% of its current population.  
    \item \textbf{Contiguity}.  Unlike most US states' requirements for Congressional districts~\parencite{congressional2021contiguity}, states do not legally mandate school attendance boundaries to be comprised of contiguous geographic units.  Still, while they exist in many districts, non-contiguous boundaries are often difficult to justify to families~\parencite{mcps2021analysis}.  We define block $b$ to be contiguous with respect to its assigned school $s$ if a line can be drawn on a map from $b$ to the block containing school $s$ without crossing through blocks zoned for any other schools.  We enforce contiguity similar to~\parencite{mehrotra1998contiguity}, with further details available in S1 of the online Supplementary Materials.  The contiguity constraint requires blocks that are contiguous with respect to their currently-zoned school must remain contiguous with respect to their zoned school under any hypothetical rezoning.  Contiguity, of course, is only a proxy for ``community cohesion'', or a desire for parents to preserve existing geographic and social networks when faced with intra-district boundary changes~\parencite{bridges2016eden}.       
\end{enumerate}

To identify plausible values for X\% and Y\% above---i.e., the travel time and school size constraints---we use the survey platform Prolific Prolific\footnote{\url{https://prolific.co/}.} to conduct a survey of 250 US-based public school parents.  We design the survey to better-understand parents' attitudes towards school diversity and the trade-offs they are willing to make to achieve more diverse schools, if any.  We gather baseline information about the parents' attitudes towards diversity, as well as information about the child's current school---including current travel times to school and average class sizes.  We then ask parents questions like the following: ``Let’s say that by changing the school zones in your district, an additional [PERCENT] of your child's classmates would come from different [CATEGORY] backgrounds.  Imagine this requires traveling further to school.  How many more minutes would you be ok with your child traveling to school in order for them to experience this increase in diversity?''.  We randomly select values for [PERCENT] and [CATEGORY] to account for different diversity scenarios (see the online Supplementary Materials for additional details).  Importantly, we acknowledge the possibility of social desirability bias in parents' responses~\parencite{pager2005bias} as an important limitation of our survey, and one that may mask several of the underlying racialized preferences for schooling described earlier.  

Acknowledging these limitations, we find that the median increase in travel times that parents would be willing to accommodate is approximately 60\% (or approximately 6 minutes, given the reported median travel time to school 10 minutes), and the median increase in class size is 15\% (or approximately 3 students, up from a reported median class size of 22).  Based on these values, we set the max travel time increase threshold to be 50\% (a conservative lower bound) and the max school size increase to be 15\%.  We do not accommodate other modes of transport, e.g. requiring students who currently walk to school to be able to continue doing so.  This may still occur under our current configurations: e.g., a student's 10 minute walk may translate into a 2 minute drive, which could increase to a max of 3 minutes under our 50\% threshold.  In the event there is such an alternative nearby option available, and the algorithm reassigns the student to it, it may still be walkable---though not guaranteed to be.  Therefore, modeling alternative commute options is an important direction for future work, especially in collaboration with school districts, who may have different transportation options and profiles.

Finally, in general, survey respondents skew more White, affluent, and suburban than national averages, with details on how respondents compare to national averages for US public schools available in S3 of the online Supplementary Materials.  These representational disparities limit the validity of the survey as a robust indicator of the preferences of families across public education systems in the US.  At best, the survey offers us a starting point for grounding our models, but one that must be refined through more participatory, community-centric efforts (a topic we return to in the discussion).

\begin{figure}
\centering
\includegraphics[width=.8\linewidth]{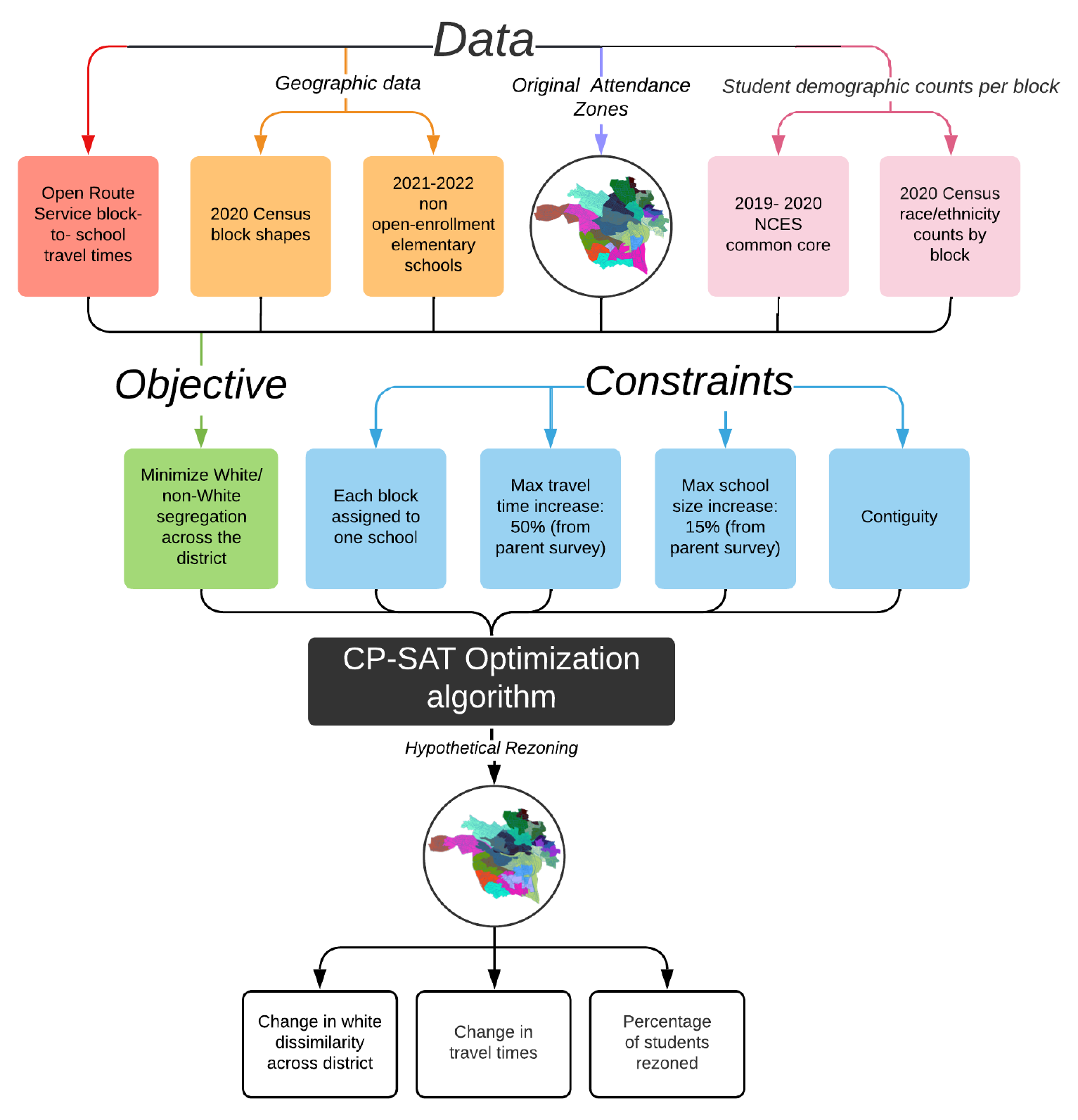}
\caption{\small{Input data, objective function, constraints, and outcome measures from our optimization model.}}
\label{fig:1}
\end{figure} 

Figure~\ref{fig:1} provides an overview of the problem setup, including the data and parameter inputs into our optimization model (with the input datasets described in more detail below), and our main outcome measures of interests: expected changes in 1) levels of segregation, 2) travel times, and 3) school switching.  Section S2 in the online Supplementary Materials contains a more detailed description of the optimization model and constraints, including our implementation of the contiguity constraint.  Given the computational intensiveness of each rezoning task, we use only one CPU core per rezoning simulation while setting a solver cutoff time of five hours and thirty minutes.  Instances are run on a parallel computing research cluster.

\subsection{Identifying districts and school attendance boundaries}
The most recent school attendance boundary survey conducted by the US Department of Education was in 2015/2016~\parencite{geverdt2018sabs}.  Therefore, for this study, we purchase 2021/2022 school attendance boundaries from the data provider ATTOM\footnote{\url{https://www.attomdata.com/data/boundaries-data/school-attendance-zone-boundaries/}.}.  Using 2020 US Census block shape files collected from the US Census website\footnote{\url{https://www.census.gov/geographies/mapping-files/time-series/geo/tiger-line-file.html}.}, we determine that a block is zoned for a particular elementary school if the centroid of that block falls within the multipolygon delineating the school's attendance zone for 3rd graders.  We exclusively use 3rd grade boundaries as our proxy for elementary schools given that 3rd grade is typically classified as an elementary grade, as opposed to e.g. 6th grade, which may be elementary or middle depending on the district/state.  In the event a district has overlapping attendance boundaries for certain schools, we map the block to the school with the smallest attendance boundary (in terms of overall area).  This occurs for approximately 7\% of blocks across the districts in our study.

We identify our sample of 98 school districts by applying the following criteria.  First, we remove districts that only have one elementary school (and hence, for which the notion of a boundary change is undefined), and those that we do not have 2019/2020 NCES school population counts for (described in the next section).  Next, for computational purposes, we include only those districts that have 200 or fewer elementary schools.  After applying these filters, we are left with 4,231 school districts in our data.  The vast majority---approximately 94\% (3,970)---have elementary schools whose boundaries are entirely ``closed-enrollment'': only those students zoned for the school can attend it\footnote{These values drop to 88\% and 86\% when considering middle (7th grade) and high (10th grade) attendance boundaries, suggesting districts are less likely to have open boundaries at younger grade levels.}.  Importantly, we note that families across even those districts with closed-boundary elementary schools may opt to attend in-district magnet programs, which do not have attendance boundaries---or opt out of the district altogether to attend an alternative (e.g., charter) school.  We discuss this possibility and its potential implications for this study further below.

The 6\% of districts excluded from our sample tend to have a slightly higher White population, slightly higher Hispanic/Latinx population, and slightly higher White/non-White segregation than the remaining 94\%.  We select the largest 100 districts (in terms of enrollment) across the 94\% of districts with closed-enrollment elementary schools.  Compared to the other 3,870 districts with no open-enrollment elementary schools, these 100 districts are (by definition) larger, but generally do not have higher levels of White/non-White segregation.  Compared to the excluded 6\%, these 100 districts are also generally larger, and \textit{do} have a higher level of White/non-White segregation.  S4 in the online Supplementary Materials offers further details on these differences.  Due to memory limitations in our computing infrastructure, we are unable to simulate alternative boundaries for two districts.  The remaining 98 districts constitute our final sample.

\subsection{Estimating students per Census block}

We use the 2019/2020 National Center for Education Statistics Common Core of Data\footnote{\url{https://nces.ed.gov/ccd/files.asp}.} to estimate the number of Black, Hispanic/Latinx, White, Native American, and Asian students at each school.  In parallel, we download 2020 Census block-level population counts for individuals who are less than 18 years of age and considered to belong to one of the above demographic groups.

With these datasets in hand, we estimate $N_{gbs}$, i.e. the number of students from group $g$ in block $b$ that attend school $s$, to be:

\begin{equation}
    \frac{C_{gb}}{C_{gB_s}} \cdot s_g
\end{equation}

Where $C_{gb}$ is the count of individuals belonging to group $g$ and living in block $b$ as estimated from the Census data; $C_{gB_s}$ is the total number of individuals from the Census data belonging to group $g$ across blocks that are zoned for school $s$ (i.e., $B_s$); and $s_g$ is the total number of students from group $g$ at $s$.  However, in cases where $s_g$ is large, we find that scaling by $\frac{C_{gb}}{C_{gB_s}}$ sometimes leads to counts per block that exceed the total number of students under 18 in that block, as defined by Census data.  Therefore, when $\frac{C_{gb}}{C_{gB_s}}$ exceeds 50\%, we replace it with $\frac{C_{b}}{C_{B_s}}$---i.e., we simply assume that the fraction of students belonging to $g$ that attend $s$ from $b$ is proportional to the fraction of total students living in $b$. Finally, we take the ceil of values and iteratively estimate counts per block (starting with the blocks with the highest value of $C_{gb}$) until all students at the school have been allocated to a home block.  This helps ensure integer student counts, and also, that the total number of students per group across all blocks is equivalent to the number attending the school per the NCES data.  Section S1 in the online Supplementary Materials contains additional details on key assumptions underlying our estimation procedure.

All Census data is collected from~\parencite{ipums2021}.  Our procedures are limited because of our inability to estimate the precise number of elementary-aged students in each block who attend their zoned elementary school, as some may attend charter, private, or within-district options with open enrollment (like magnet programs).  Even though certain demographic groups disproportionately may exercise school choice in different settings~\parencite{schachner2022LA,bischoff2020imbalance,rich2021seg}, because our estimates are based on ground truth school enrollments by demographic group, this differential uptake of school choice is likely to bias our block-level estimates only if families who are part of the same demographic group and assigned to the same school have different rates of school choice uptake that are correlated with the block in which they live.  It is not immediately obvious why this might happen, but there are certainly possible explanations (for example, a particular block might house a popular charter or other alternative school option).  This could affect the results of our boundary redrawing by either over or understating how much alternative boundaries might impact school diversity.  For example, districts with a high fraction of non-White students that have disproportionate numbers of White families opting out of zoned schools in certain blocks compared to others may overstate how much alternative boundaries could increase integration; conversely, higher fractions of non-White families opting out across these blocks (e.g., due to charter options with lotteries that reserve seats for different demographic groups) may understate it.

In practice, districts cannot know exactly which students living within their boundaries opt for charter schools or other out-of-district options.  However, they \textit{can} know the locations of students attending within-district schools---and so, through future collaborations with districts, we can much more accurately determine student counts per blocks.  School choice, however, not only impacts block-level student counts (model inputs); it also impacts how likely families are to adhere to new boundaries (model outputs), and hence, the extent to which diversity and integration objectives are actually achieved post-rezoning.  In an ideal world, we would have access to a clairvoyant capable of perfectly anticipating which students are likely to exercise school choice in the face of proposed student assignment policy changes \textit{before} these policies are implemented.  However, modeling and anticipating family demand for schools continues to be an open research problem~\parencite{pathak2021demand} and challenging practical task for districts.  In the results section, we present simple opt-out scenarios based on charter and magnet choice patterns across districts as a first step towards illustrating how school choice might impact the boundary-based integration strategies and outcomes we focus on in this study.

\subsection{Estimating travel times}

We use the OpenRouteService API~\parencite{ors2022} to estimate travel (driving) times between block centroids and schools in each district.  Given the large number of travel times to compute (millions in some of the larger districts) and the publicly-hosted API's rate limits, we compile and run a local instance of the API on our own server, which enables us to submit an arbitrary number of queries.  Queries are comprised of latitude/longitude pairs for a starting location (Census block centroid) and ending location (school location).  Travel times do not account for traffic patterns.  

\section{Results}

\begin{figure}
\centering
\includegraphics[width=\linewidth]{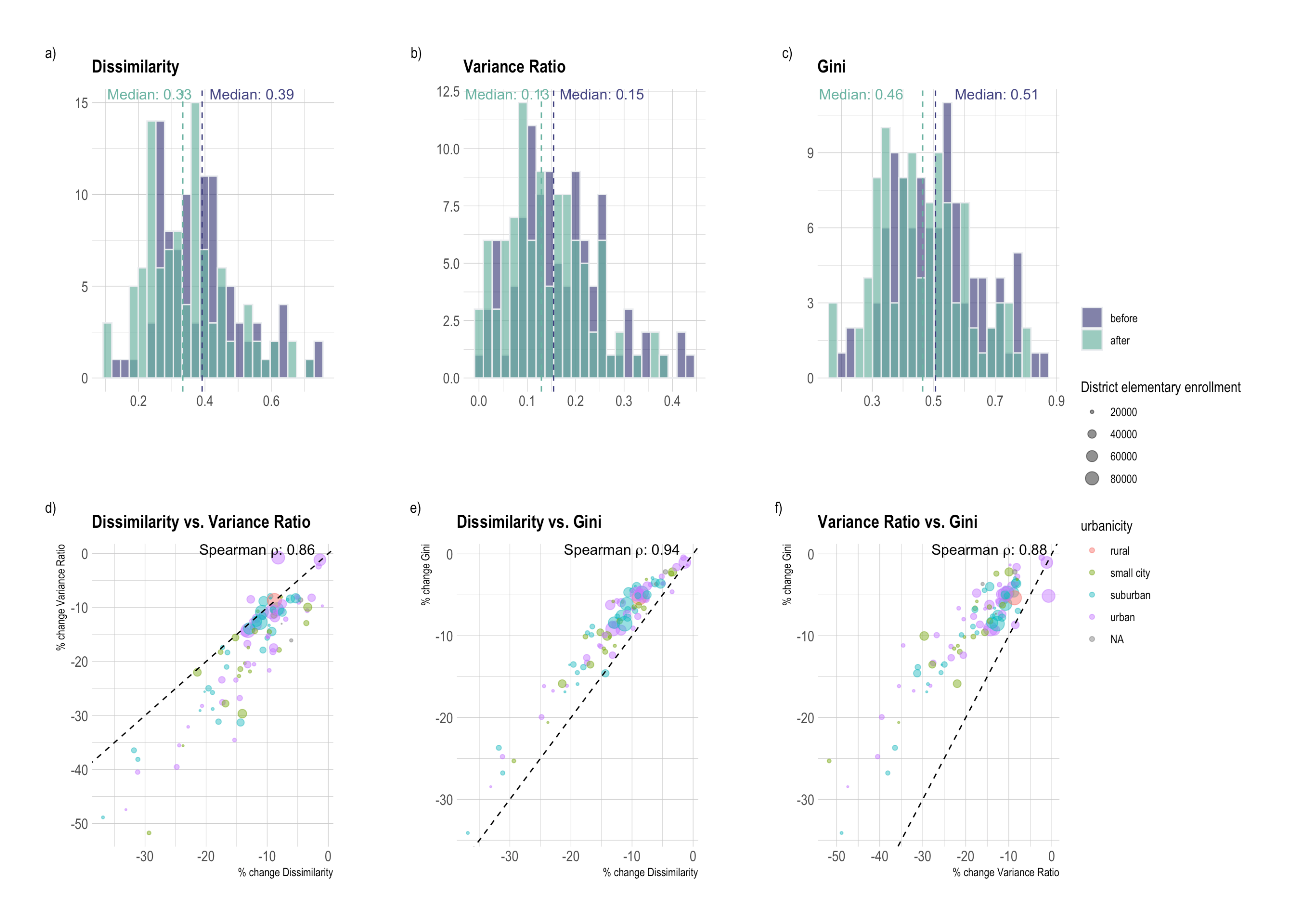}
\caption{Estimated impacts of attendance boundary changes on levels of segregation across districts.  Plots \textbf{a)-c)} depict the before and after-rezoning distributions of segregation scores according to the Dissimilarity, Variance Ratio, and Gini indices for the 98 districts in our sample.  In plots \textbf{d)-f)}, each bubble represents a district, and the plots show the strong Spearman rank correlations between the relative reductions in segregation across districts as measured according to our three indices. 
 In general, the relative changes described by each measure of segregation are quite similar across districts.}
\label{fig:2}
\end{figure} 

We begin by analyzing White/non-White segregation scores for our 98 districts across the Dissimilarity ($D$), Gini ($G$), and Variance Ratio ($V$) indices described earlier.  Figures~\ref{fig:2}(a-c) illustrate the distribution of these segregation scores across the districts in our sample before and after producing our hypothetical rezonings.  The median values of $D$, $G$, and $V$ before rezoning are 0.39, 0.51, and 0.15, respectively.  Following rezoning, the median across these metrics decreases to 0.33, 0.46, and 0.13, respectively---corresponding to a median 12\%, 7\%, and 14\% relative decrease when computing pairwise changes per district.  The post-rezoning scores for $D$ and $V$ are based on simulations that seek to directly minimize these values as the core objective functions.  While we also simulated rezonings designed to directly minimize $G$, the $|S|^2$ number of terms in the objective function (where $S$=number of schools) added significant computational complexity, limiting both the number of districts we could simulate changes for (only 91 out of 98 districts) and the quality of rezonings produced by these simulations.  Given the similarities between how $D$ and $G$ measure segregation, the results for $G$ presented here are computed across simulations optimizing for $D$---which, surprisingly, produce a larger median relative decrease in $G$ across districts (7\%) than those simulations directly optimizing for $G$ (5\%), further underscoring potential performance and solution quality issues.

Figures~\ref{fig:2}(d-f) illustrate Spearman rank correlations ($\rho$) between the relative reductions in segregation across districts according to our three metrics.  The correlation between relative reductions in $D$ and $V$ across districts is $\rho=0.86$; between $D$ and $G$ is $\rho=0.94$; and between $V$ and $G$ is $\rho=0.88$ ($p << .001$ in all cases). In general, it appears that regardless of these different definitions of segregation, our rezoning algorithms produce similar relative reductions in segregation across the districts in our sample.  Therefore, for simplicity throughout the remainder of the paper, we select one of these metrics as the basis of further investigating the results of our models.  In particular, we select $V$ because it both respects the transfers/exchanges principal and also offers insight into how segregation is actually experienced within schools (on average) by students from different racial and ethnic backgrounds.  We note $V$'s lower score relative to $D$ and $G$, which we hypothesize may be due to the fact that $V$ also measures exposure---and so, districts that have a larger fraction of non-White students may increase the chances that White students are exposed to them in schools.  Indeed, computing a Spearman correlation between $D - V$ per district (before rezoning) and the percentage of non-white students across districts yields $\rho=0.74, p << 0.001$ (with similar results for $G - V$).  The larger the fraction of non-White students in a district, the less the Variance Ratio index's value aligns with the values of evenness-based measures of segregation like $D$ and $G$, likely because more non-White students in a district increases the chances that the average White student encounters them at school.  Even with this trend, though, there is noticeable segregation across districts according to $V$, and hence, an opportunity to explore how much boundary changes might reduce it.

\begin{figure}
\centering
\includegraphics[width=\linewidth]{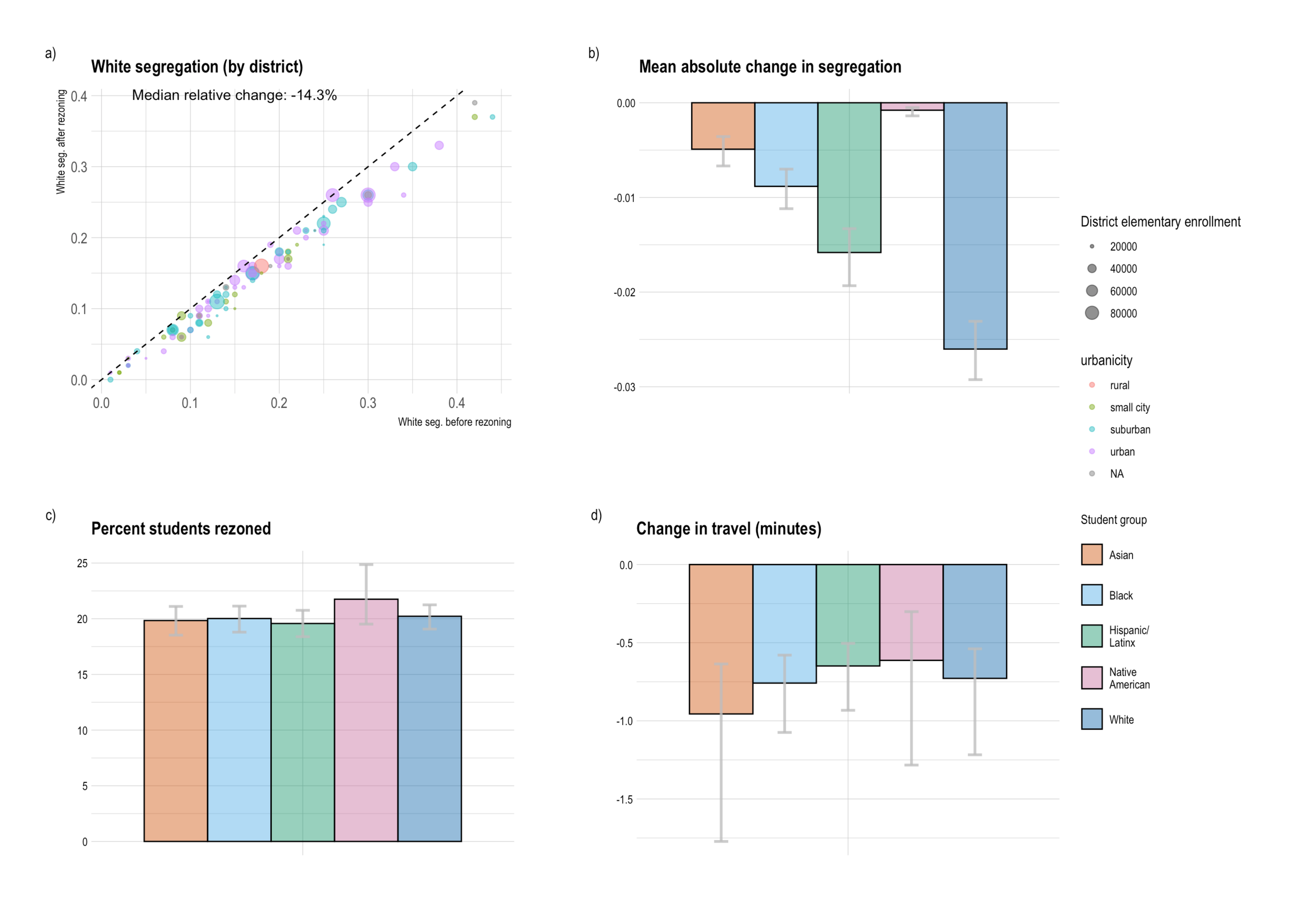}
\caption{Results from our rezoning algorithm. \textbf{a)} illustrates pairwise changes in $V$ across districts; \textbf{b)} shows that segregation scores for all racial/ethnic groups decrease, albeit marginally, under the proposed rezonings. \textbf{c)} illustrates that, on average across districts, school switching under the depicted rezonings are relatively evenly distributed across racial and ethnic groups. Finally, \textbf{d)} illustrates that the depicted boundary changes might actually slightly decrease average travel times across school districts and demographic groups. Together, these findings show that there are pathways to more integrated schools across districts that may not require large sacrifices by families.}
\label{fig:3}
\end{figure} 

Figure~\ref{fig:3} offers a deeper look at changes in $V$, school assignments, and travel times produced by our models across districts (error bars depict 95\% bias-corrected and accelerated confidence intervals, computed using the boot library in R~\parencite{r2021boot, davison1997boot}).  From~\ref{fig:3}(a), we observe the aforementioned 14\% median relative decrease in $V$ indices across districts after our hypothetical rezoning.  Districts in our sample range in their values for $V$: the most segregated district has a value of $V=0.44$, while the least segregated has a value of $V=0.01$.  Conducting exploratory correlational analyses, we observe no statistically-significant association between a relative decreases in $V$ and urbanicity ($ANOVA~F=0.66, p=0.58$), and only moderate relationships between relative decreases in $V$ with enrollment size ($Spearman~\rho=0.35, p < 0.001$) and initial levels of $V$ ($Spearman~\rho=0.35, p < 0.001$).  These results suggest that larger and more segregated districts have more scope for intra-district attendance boundary changes to increase integration, but given the relatively moderate associations with these variables, that the nuanced geographic and demographic contexts of each district is likely to play an important role in how much such boundary changes can foster more diverse and integrated schools.

Figures~\ref{fig:3}(b)-(d) illustrate the potential costs of achieving these reductions in segregation.  From (b), we see that reducing White/non-White segregation would not lead to higher segregation levels for other racial groups (i.e., Black/non-Black; Hispanic/Latinx non-Hispanic/Latinx; etc).  In fact, the other racial groups would also experience reductions in segregation under the depicted rezonings.  In (d), we see that, on average, approximately 20\% of students from different groups would be required to switch schools, and that the burden of school switching could be distributed approximately evenly across student groups.  While 20\% represents a relatively large fraction of students, it is less than the nearly 40\% of parents in our survey who expressed a willingness to switch schools if their district redrew attendance boundaries.  From an implementation perspective, districts may also phase boundary changes in gradually instead of all at once, reducing the number of students required to switch schools in any given year.  The literature on the impact of school switching on student academic and subjective well being outcomes is mixed, with some findings illustrating positive benefits conditional on switching to attend better schools, and others illustrating adverse consequences~\parencite{hanushek2004switching,schwartz2017switching}.  Weighing the potential disruption costs of school switching alongside the potential gains of more integrated schools is important when determining when and how to make boundary changes.

Somewhat surprisingly, plot (e) shows that average school switcher would actually experience a \textit{decrease} in their travel times to and from school, despite the fact that our model permitted up to a 50\% increase in travel times for any given family.  This is notable because it suggests that 1) long-range ``busing''~\parencite{delmont2016busing} is not necessarily required to achieve more diversity in schools, and 2) some existing attendance boundaries may potentially be drawn (``gerrymandered'') in ways that assign students to schools further from their homes, resulting in slightly higher levels of segregation as a result~\parencite{richards2014gerrymandering}.  We note the speculative nature of this latter point, especially given the existence of research suggesting that irregulary-shaped boundaries may actually contribute to \textit{greater} integration~\parencite{saporito2016shapes}.  Indeed, it is possible families may have moved after the implementation of such boundaries precisely to avoid more integrated schools, producing a net increase in segregation.  Further research is needed to better understand precisely why it appears that current boundaries could be redrawn to foster integration while also reducing travel times.  Finally, an important observation from Figures~\ref{fig:3}(c) and (d) is that, again on average across districts, the potential costs of desegregation are fairly distributed across the depicted racial and ethnic groups.

\subsection{Sensitivity analyses}

\begin{figure}
\centering
\includegraphics[width=\linewidth]{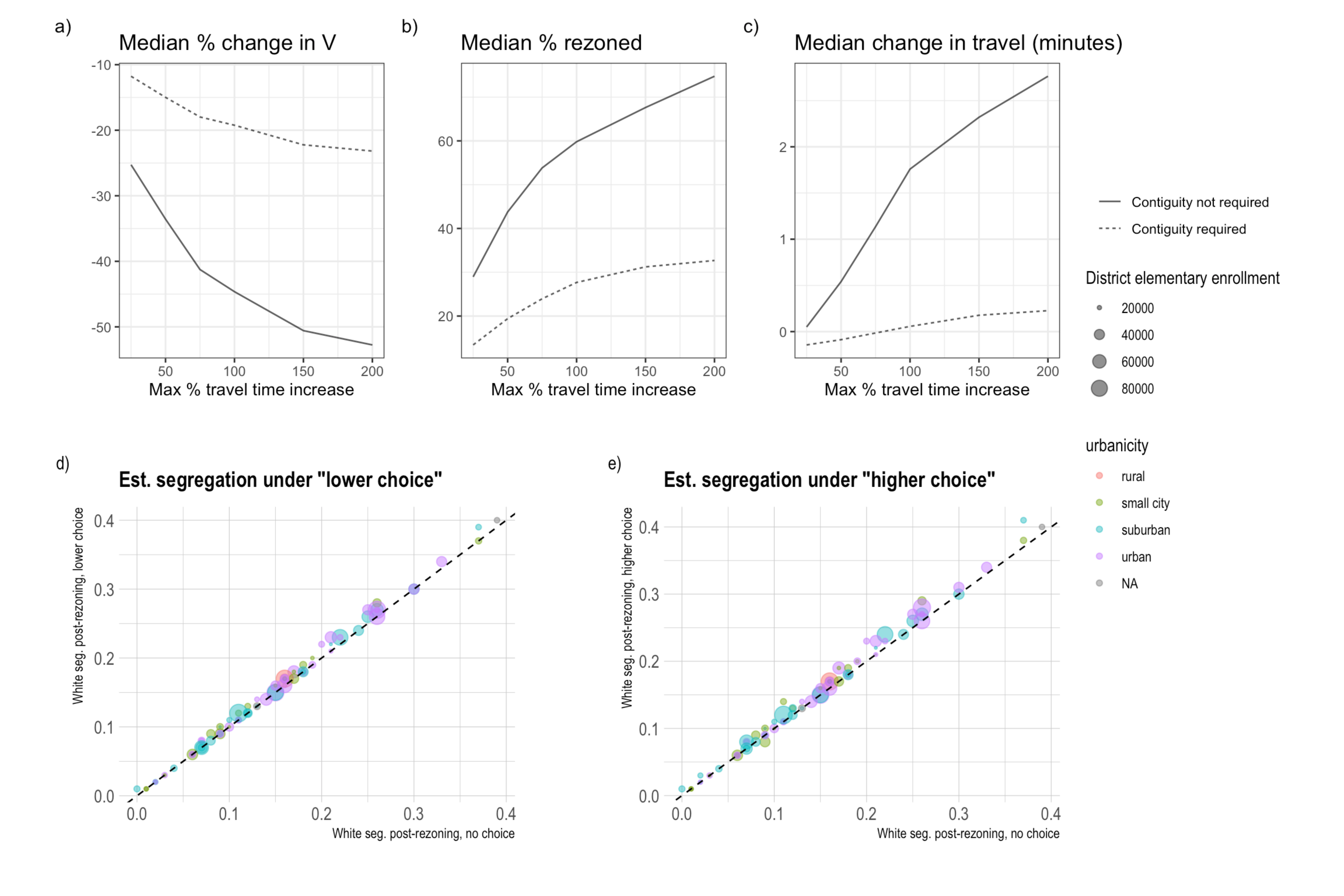}
\caption{Sensitivity analyses of main results.  Plots \textbf{a)-c)} depict how much segregation, percentage of students who are rezoned, and travel times for rezoned students might change as a result of changing travel time and contiguity constraints (while fixing max school size increases to 15\%).  Values are medians across the 98 districts in our sample.  In general, dropping contiguity constraints appears to have the largest potential impact on reducing segregation, but we estimate would also produce much more school switching and higher travel times for students.  Plots \textbf{d)} and \textbf{e)} illustrate simple models for how much demographic group-specific patterns of choice might impact reductions in segregation produced by boundary changes.  In both of these cases, we estimate patterns of choice would lead to a median relative reduction of approximately 10\% across districts in our sample, down from the 14\% indicated by our models without modeling choice-based opt-outs.}
\label{fig:4}
\end{figure} 

A median 14\% relative decrease in segregation across districts represents a non-trivial step towards more integrated schools, yet on its own is quite modest and far from achieving full integration.  It highlights how, under our selected constraint values, there are inherent limits to how ``sticky'' the issue of segregation is, and the limited extent to which intra-district boundary changes might promote more diverse and integrated schools.  To explore how segregation might change under different parameter configurations, Figures~\ref{fig:4}(a-c) fixes maximum school size increases to 15\% and illustrates how changing travel time and contiguity constraints might impact median levels of segregation, school switching, and travel times for rezoned students across our 98 districts.  For example, setting the max travel increase threshold to 100\% (or in the most extreme case, allowing families to experience a doubling in travel time to school) could yield a median relative decrease in segregation of nearly 20\%, but would require a median of nearly 30\% of students to switch schools and experience a slight average increase in travel times (though these still appear to be marginal).  Keeping the travel time increase at a maximum of 50\% but dropping the contiguity constraint could yield a median relative decrease in segregation of 35\%, but would require nearly 45\% of students to switch schools, and experience a half-minute average increase in travel to school.  Applying such relaxations together---not requiring contiguity and allowing even larger increases in travel times of a miximum of 200\%---could decrease segregation by over 50\%, but would also require a median of over 70\% of students to switch schools and experience a median travel time increase of nearly 3 minutes.  Perhaps most salient in these plots is that requiring contiguity appears to significantly limit the extent to which intra-district boundary changes might foster more diverse and integrated schools.

A critical limitation of all of our results so far is that they do not factor in the likelihood of complex system dynamics that could manifest if districts actually \textit{did} adopt the rezonings described here---for example, neighborhood relocation (e.g. ``white flight'') in response to unfavorable rezonings~\parencite{reber2005flight,nielsen2020denmark}, or the disproportionate use of school choice by families to opt for other district or charter options that enable them to circumvent the effects of changing boundaries.  All of these could affect the extent to which the methods proposed thus far can actually help districts achieve greater racial and ethnic integration in schools.  Unfortunately, as discussed earlier, anticipating rates of family opt-out is a challenging task. We explore a simple model of family opt-out based on existing patterns of charter and magnet school choice within the districts in our sample to offer a preliminary investigation of how much school choice might undermine boundary-based student assignment policy changes seeking to foster more diverse and integrated schools.  We use the NCES Common Core of Data to identify elementary charter and in-district magnet schools located within at least one of the elementary school attendance boundaries across the 98 districts in our sample.  Of these 98 districts, 84 contain at least one charter or magnet school, totaling 988 total such schools.  We estimate that the median ratio of elementary students per district attending a charter or magnet school compared to a closed-enrollment boundary-based school is approximately 12.8\%.  There is, however, high variance across districts: the ratio in the district where choice is most prevalent is 67\%; in the least-prevalent district, it is less than 1\%.  In 26 out of 84 districts, White students are more likely to attend choice programs than non-White students, suggesting that different public school choice contexts may foster different choice patterns among families from different demographic backgrounds~\cite{bischoff2020imbalance, schachner2022LA}.

For each district, we compute the ratio of elementary students per racial/ethnic group (White, Black, Hispanic/Latinx, Asian, Native American) who attend a charter or magnet school compared to a closed-enrollment boundary school.  These estimates serve as the basis of two ``opt-out'' scenarios we model to explore how much the prevalence of choice across districts might undermine boundary-based policy changes seeking to foster greater integration.  In one scenario---``lower choice''---we assume that students who are rezoned under a hypothetical boundary change opt-out at a rate of $\frac{1}{2}$ their existing demographic group-specific choice uptake ratio.  To make this concrete: if our algorithm re-assigns a census block with 20 Black students, and district-wide, the ratio of Black students attending a charter or magnet school located in the district compared to a closed-enrollment boundary-assigned school is 10\%, then we reduce the count of Black students who actually transition to the new school by $\frac{1}{2}\cdot 0.1 \cdot 20$, or one student.  The second scenario---``higher choice''---works similarly, but simply applies the district-wide choice ratio: in the earlier example, $0.1\cdot 20$, or two Black students, would opt out.  The key assumption motivating these two scenarios is that the rate of existing public school choice in a particular district is likely to influence the extent to which families have and exercise options beyond their boundary-assigned school.  Of course, actual rates of boundary-assigned school opt-out will vary not only as a function of existing choice options, but also, where those schools are located relative to families; the programs they offer; the specific nature of how a rezoning might alter the features of their boundary-assigned school; and several other factors.  Indeed, exploring more robust models to anticipate such choice in the face of potential boundary changes is important avenue for future work. 

Figures~\ref{fig:4}(d-e) show how much we expect levels of $V$ to change (compared to no opt-out) across both of these scenarios.  In both cases, $V$ decreases by a median of approximately 10\%, down from 14\% in the original case where there are no expected opt-outs.  This suggests that, at least according to these simplistic models, choice patterns might undermine integration outcomes by a factor of nearly one third.  Crucially, these scenarios do not account for potential private school opt-outs, capacity constraints at choice-based schools, or other real-world complexities, and thus should be viewed as simplistic early efforts to illustrate how the dynamics of family preferences and school selection might impact integration objectives that districts might pursue through boundary changes. 

\subsection{District case studies}
Thus far, we have discussed average expected impacts of boundary changes in integration and other outcomes.  Yet these averages likely mask important heterogeneities across different types of districts.  To explore some of these heterogeneities, we conduct two case studies.  The first involves the most segregated district in our sample, Atlanta Public Schools, which has a White/non-White Variance Ratio index of 0.44 and serves nearly 23,000 students across 44 closed-enrollment attendance boundary elementary schools.  The second involves the district closest to the median level of segregation across districts in our sample: Garden Grove Unified District, California, which has a Variance Ratio index of 0.15 and serves over 20,000 students across 47 closed-enrollment elementary schools.

\begin{figure}
\centering
\includegraphics[width=\linewidth]{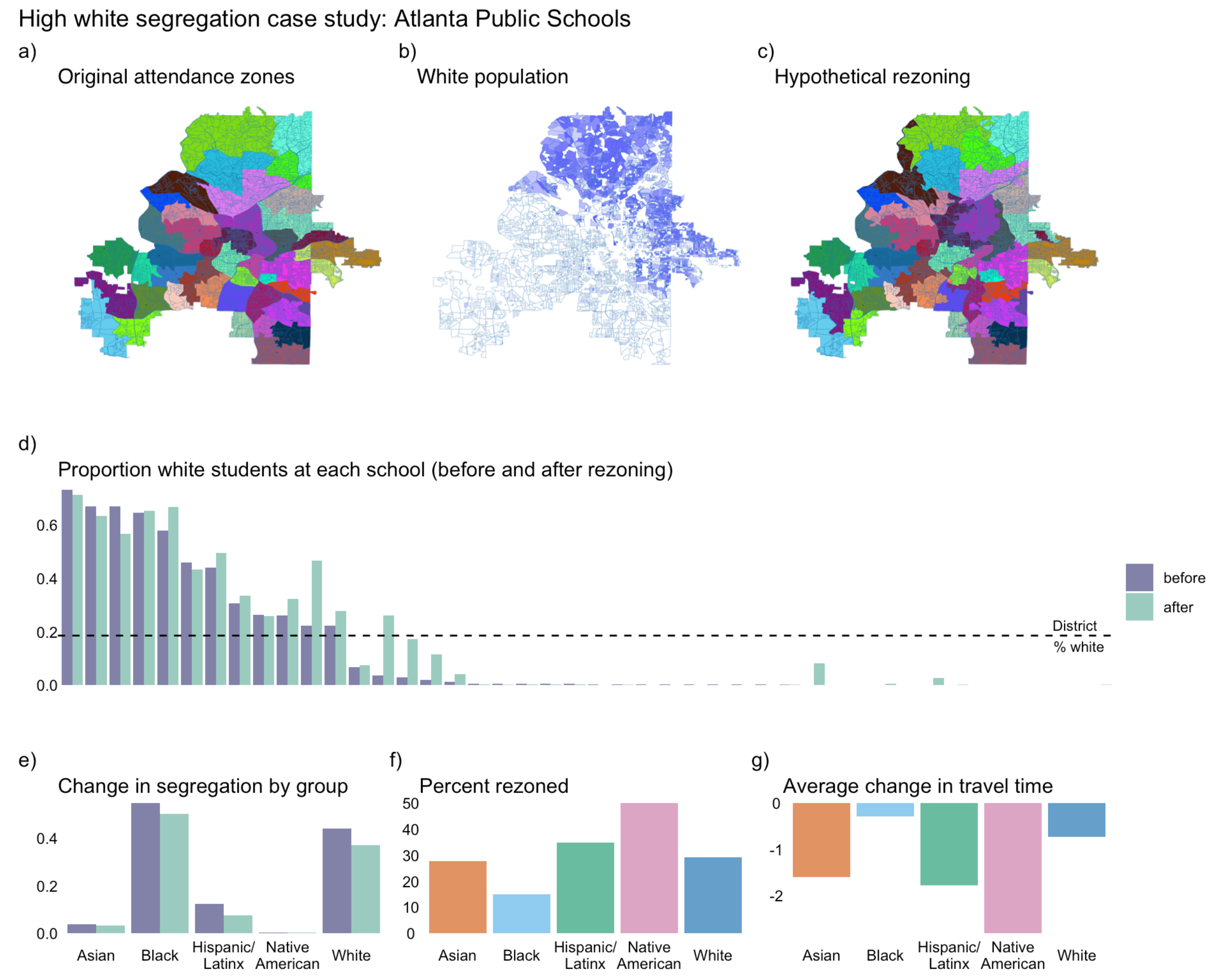}
\caption{\tiny{Case study for the most White/non-White segregated district in our sample, Atlanta Public Schools, which has a segregation score of $V=0.44$.  The shapes in \textbf{a)} through \textbf{c)} represent 2020 US census blocks; the colors in \textbf{a)} delineate 2021/2022 school attendance boundaries for the depicted district ("status quo"); \textbf{b)} shows the estimated percent of each block's population estimated to be white (darker blue implies a higher percentage), with blocks removed if they are estimated to have a student population of zero. \textbf{c)} shows the rezoning produced by our algorithm.  \textbf{d)} shows the expected change in the proportion of students at each elementary school in the district who are White, before and after rezoning.  \textbf{e)} through \textbf{g)} show the anticipated changes in segregation scores, percentage of students needing to be rezoned, and change in travel times for each demographic group.  The results reveal several notable findings.  First, as expected, the most dramatic boundary changes appear to occur in school boundaries that fall at the interface of White and non-White parts of the city.  Additionally, as shown in \textbf{d)}, changes in school-level distributions tend towards the district White student percentage, with a handful of schools experiencing the most dramatic changes.  However, the share of White students is also \textit{increased} at several schools that already have a White share higher than the district, illustrating trade-offs district leaders may be faced with when deciding which schools to target with desegregation efforts.  From \textbf{e)} and \textbf{g)}, we see that all student groups would experience reductions in segregation and travel times, respectively, but \textbf{f)} shows disparities in which students might be rezoned---with the largest fraction found among Native American (8 out of 16), Hispanic/Latinx (613 out of 1,756), and Asian (93 out of 336) students.}}
\label{fig:5}
\end{figure} 

\begin{figure}
\centering
\includegraphics[width=\linewidth]{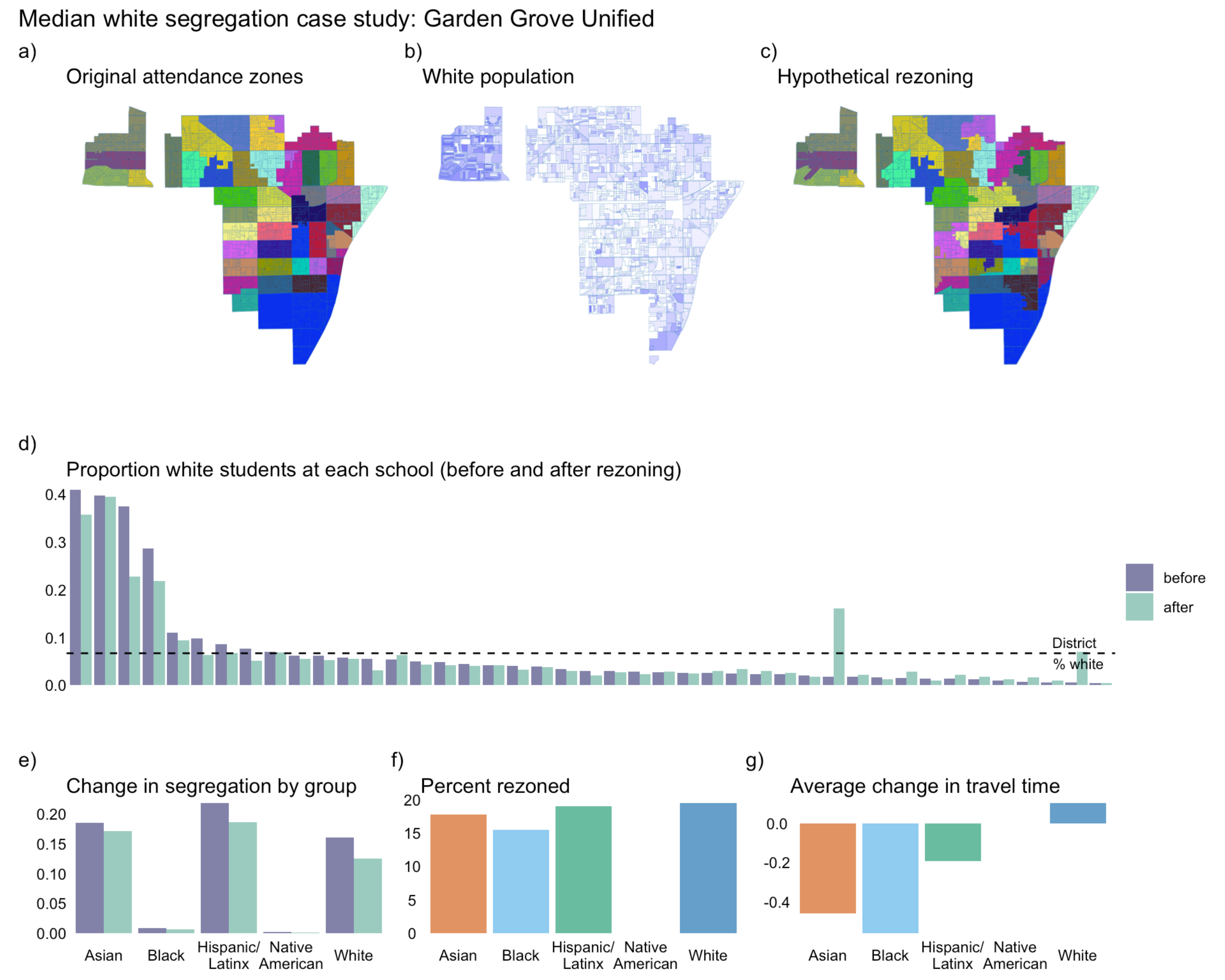}
\caption{\small{Case study for a typical (median) White/non-White segregated district in our sample, Garden Grove Unified, CA, which has a segregation score of $V=0.15$.  Plots \textbf{a)}-\textbf{g)} defined analogously to Figure 4.  In general, we observe similar trends as the Atlanta case study, with fewer disparities in the percentage of students across demographic groups who would be rezoned, and a marginal increase in travel times for White students in the district who would be rezoned under the depicted boundary change scenario.}}
\label{fig:6}
\end{figure} 

Figures~\ref{fig:5} and~\ref{fig:6} illustrate the outputs of these case studies.  Maps (a)-(c) in each figure illustrate the present-day elementary school attendance boundaries, the relative prevalence of White students per Census block, and our hypothetical rezoning.  Examining (d) illustrates how the fraction of students in the depicted group would change at each school after implementing the alternative zoning, compared to before.  As expected, rezonings generally move school-level demographics to reflect district-level proportions as much as possible, sometimes shifting the percentage of White students at a given school by several fold, as seen in Figure~\ref{fig:5}(d).  The most substantial changes also appear to occur within a small subset of schools, with most schools across the district experiencing little or no change.  However, some schools become more segregated with respect to the district: that is, a subset of schools which already have a proportion of White or non-White students exceeding district-level proportions see an increase in their White or non-White share, respectively, and therefore diverge further from, instead of converging to, district-level shares of White/non-White students.  These fluctuations are more extreme in Atlanta compared to Garden Grove Unified, and to be expected given our Variance Ratio index objective function, which optimizes an aggregate district-level measure without explicitly requiring that reductions in segregation be evenly redistributed across individual schools.  In practice, the choice of objective function, overall desegregation goals (including specific schools of focus), and even notions of fairness are likely to be context-specific and require input and domain expertise from both districts leaders and families.  While our measures of segregation here are defined at the district level, we return to a point made in the introduction---that there will be inherent limits to how much racial balancing at schools can be achieved through intra-district changes alone.  For example, according to our sample, Atlanta Public Schools has nearly 23,000 students at boundary-based elementary schools, approximately 4,200 of whom are White.  Cobb County, adjacent to Atlanta, has approximately 48,000 students in our sample, nearly 18,000 of whom are White.  Exploring inter-district changes between districts with stark racial and ethnic gradients such as these, while more politically challenging, may offer even more hope for fostering more diverse and integrated schools. 

Plots (e) through (g) in each case study depict the changes in segregation, school assignments, and travel times expected across demographic groups.  In both cases, we see slight reductions in segregation and travel times across demographic groups, with the exception of White students in the Garden Grove Unified case, who would experience a marginal ($<<1$ minute) increase in travel each way.  Furthermore, in Figure~\ref{fig:6}(f) (Garden Grove Unified), we see a relatively balanced school switching requirement across demographic groups\footnote{With the exception of Native American Students, who constitute less than 0.1\% of the students across the district elementary schools included in our sample}; however, in the corresponding plot for Atlanta (Figure~\ref{fig:5}(f)), we see larger disparities across groups.  Most notably, White, Hispanic/Latinx, and Asian students are 2-3x as likely as Black students to have to switch schools; Native American students are nearly 5x as likely.  Notably, however, Native American, Asian, and Hispanic/Latinx students constitute a small fraction of the overall student body in Atlanta's elementary schools.  These differences show that even though in an aggregate sense across districts, school switching is fairly distributed across groups, these results will likely vary by district.  This highlights both several of the trade-offs district officials might need to make in order to achieve more integrated schools, but also, opportunities for more creative approaches to modeling, constraining, and ultimately addressing the issue of changing policies in order to mitigate segregation.

\section{Discussion}

Our results demonstrate that there are practical and fair pathways to changing attendance boundaries in order to achieve more diverse schools, though the impact these policies have on individual schools and different student groups---for example, who would be required to switch schools, and how much school switching might either further or hinder these students' academic progress---can vary across districts.  This reality calls for a nuanced and district-specific approach to modeling, evaluating, and eventually adopting potential boundary changes.  Particularly notable is that there exist alternative boundary scenarios that might reduce segregation \textit{and travel times} across districts, highlighting that these are not always at odds, and contrasting with the public narrative around long-range busing that emerged among many majority-race members of the population during desegregation efforts in the 20th century~\parencite{delmont2016busing}.  

While we observe a median relative decrease of 14\% in the White/non-White Variance Ratio index of segregation across districts after our hypothetical rezonings, these improvements are largely a function of the constraints our models impose, and still constitute only small steps towards addressing issues of White/non-White segregation.  Nevertheless, as our sensitivity analyses show, changing constraint values (and particularly, dropping contiguity) can have a multiplier effect on how much alternative boundaries might reduce segregation.  On the other hand, our preliminary modeling suggests that family opt-outs from boundary-assigned schools in favor of charter or magnet programs may slightly undermine integration objectives.  To support explorations of these sensitivities and the impact different policies might have on individual schools and demographic groups, we release a public dashboard\footnote{\url{https://www.schooldiversity.org/}.} and its underlying code and data illustrating different boundary scenarios and outcomes for the districts we explore in this study.  We invite researchers to use these resources to explore new models that capture more of the nuances and specifics individual districts often consider when making boundary changes (some of which we discuss below).  We also invite districts and families to explore the outputs in the dashboard and comment on their viability as starting points for informing realistic policies for fostering more diverse schools.

Parents' racialized preferences for where they live and send their children to school will continue to act as formidable headwinds challenging even the most thoughtful and well-designed efforts to foster more diverse schools.  Our study does not contribute to answering the normative question of how to change these preferences, or the political one of whether school districts can garner the will to implement policies that improve integration.  However, it offers an empirical contribution that we believe may be of interest to both researchers and school districts: that such improvements appear to be possible across many districts, and that they can be achieved with practical and fair tradeoffs.  Even then, which tradeoffs count as "practical" and "fair" will differ across communities and individuals, and across racial/ethnic and class lines.  This points to a number of limitations in our current study, which in turn open the door to new and exciting directions for future work.  We classify these limitations as opportunities across three interconnected categories that we invite researchers and interested practitioners to explore in greater detail: data, model, and broader relevance to education policy efforts.

With respect to data, our method relies on estimated counts of students per group, per Census block---which could be improved by obtaining ground-truth data through district collaborations.  We are also unable to factor in other data, like transportation costs, that districts might weigh as they decide on rezoning policies.  The limited window that Free/Reduced-priced lunch provides into students' socioeconomic status~\parencite{harwell2010frl}, coupled with the limited availability of family socioeconomic indicators at the Census block level, prevents us from exploring socioeconomic dimensions of segregation, which several districts seek to mitigate~\parencite{potter2016stamford}.  Working closely with specific district partners to obtain and incorporate more detailed, historical, and up-to-date data may help alleviate many of these issues. Finally, we proxy ``community cohesion'' with contiguity.  In reality, a family's community, and students' friends, are a function of geography along with many other (potentially unobservable) factors.  Developing more nuanced ways of determining and factoring in notions of community into rezoning models may open the door to new boundary configurations that relax the constraining impact of contiguity constraints while still satisfying family and district-level preferences.  

There are also a number of model improvements that may make our results more useful in practice.  Balancing utilization across schools, limiting the percentage of students who are rezoned (a la ~\parencite{mcps2021analysis}), and even more explicitly factoring in fairness requirements instead of merely analyzing fairness post hoc serve as important directions for modeling improvements.  To ensure increases in elementary school diversity also propagate to middle and high schools, exploring objective functions that factor in feeder patterns and account for the full K12 lifecycle---instead of only the earlier years---may also produce more practical and desirable boundary changes.  As mentioned earlier, we may also expand our model to incorporate historical data or domain expertise to predict how a given rezoning might spark families to leave neighborhoods and/or disproportionately leverage school choice to access other district or charter options---and factor these possibilities into the optimization process.  Finally, given that diversity may not today be a core consideration or impetus for redrawing boundaries in most districts, we might augment our models to aid district policymakers in simulating new boundaries when exploring questions more germane to their day-to-day, like determining locations for new schools, or deciding which schools to shut down (for example, in response to declining enrollment).  With minor extensions, the models we present here can aid with these decisions while still foregrounding their potential impacts on diversity, travel, and other outcomes of interest.

Perhaps the biggest open question from our study is: how might families and district leaders respond to these hypothetical rezonings, and how much could they actually increase diversity in schools?  We believe this is an important avenue for follow-on research, and a critical part of translating this research into education policies that help promote school diversity.  As discussed throughout the study, boundary design matters, but so does its interplay with how families opt for choice-based schools (should they exist in their districts).  Therefore, it is critical to explore how the methods presented here may more accurately anticipate family opt-out rates in favor of charter or magnet programs---or even private schools; how they might inform the design of ``zones of choice'' that define meta-boundaries for clusters of schools that parents can then choose amongst~\parencite{campos2022zones,allman2021sfusd}; and several other student assignment and school choice policies, many of which are emergent.  Furthermore, as discussed in the introduction, most segregation in schools can be attributed to boundary delineations \textit{between} districts, not simply those within them.  Expanding geographic scope to explore between-district boundary changes (a more challenging computational problem as well) may help yield policy simulations that produce more practical and effective pathways to integrated schools.  Finally, capturing and factoring in input from both families and district leaders, for example, through participatory~\parencite{kensing1998pd} and value sensitive~\parencite{friedman2013vsd} design methods may further help inform school desegregation policies that are realistic and practically-achievable.

Changing school demographics does not guarantee more diverse friendships, sharing of social capital and resources, greater empathy for different life experiences, and other potential gains that can ultimately benefit all students~\parencite{moody2001tracking, tatum1997cafeteria, chetty2022socialcapitalII}.  Yet it is a necessary first step towards achieving many of these downstream outcomes.  We hope this study is a useful building block to support future work on this critical topic.

\section{Acknowledgements and funding sources}

Funding for this project was provided by the MIT Center for Constructive Communication.  The authors declare no competing interests.  The research was approved by MIT's Institutional Review Board.  We thank Peter Bergman, Jennifer Candipan, Kumar Chandra, Tyler McDaniel, Tomas Monarrez, Laurent Perron, Halley Potter, Francine Stephens, Nathanial Thomas, participants at Harvard's Urban Data Lab, and this paper's anonymous reviewers for their helpful comments and guidance.    

\printbibliography

@article{
    reardon2018testgaps, 
    title="{The Geography of Racial/Ethnic Test Score Gaps}",
    author={Reardon, Sean F. and Kalogrides, D. and Shores, K.},
    journal={Stanford Center for Education Policy Analysis, Working Paper No. 16-10}, 
    year={2018}
}

@misc{ors2022,
  author = {GIScience},
  title = {Open Route Service},
  year = {2022},
  publisher = {GitHub},
  journal = {GitHub repository},
  url = {https://github.com/GIScience/openrouteservice},
}

@misc{ipums2021,
  author = {Manson, Steven and Schroeder, Jonathan and Riper, David Van and Kugler, Tracy and Ruggles, Steven},
  title = {IPUMS National Historical Geographic Information System: Version 16.0 [dataset]},
  year = {2021},
  publisher = {Minneapolis, MN: IPUMS},
  doi = {http://doi.org/10.18128/D050.V16.0},
}

@article{gurnee2021fairmandering,
  title = "{Fairmandering: A column generation heuristic for fairness-optimized political districting}",
  author = {Gurnee, Wes and Shmoys, David B.},
  journal = {arXiv: 2103.11469},
  year = {2021}
}

@misc{brownvboard,
  title = {Brown v. Board of Education of Topeka (1)},
  % year = {n.d.},
  howpublished = {Oyez},
  url = {https://www.oyez.org/cases/1940-1955/347us483},
  urldate = {2022-02-07}
}

@misc{potter2016stamford,
  author = {Potter, Halley},
  title = {Stamford Public Schools: From Desegregated Schools to Integrated Classrooms},
  year = {2016},
  howpublished = {The Century Foundation},
  url = {https://tcf.org/content/report/stamford-public-schools/},
  urldate = {2022-02-07}
}

@article{card2016gt,
  title="{Universal screening increases the representation of low-income and minority students in gifted education}",
  author={Card, David and Giuliano, Laura},
  journal={Proceedings of the National Academy of Sciences},
  volume={113},
  number={48},
  pages={13678--13683},
  year={2016}
}

@article{comer1988integration,
  title="{Educating Poor Minority Children}",
  author={Comer, James P.},
  journal={Scientific American},
  volume={259},
  number={5},
  pages={42--49},
  year={1988}
}

@article{billings2013cms,
  title="{School Segregation, Educational Attainment, and Crime: Evidence from the End of Busing in Charlotte-Mecklenburg}",
  author={Billings, Stephen B. and Deming, David J. and Rockoff, Jonah},
  journal={The Quarterly Journal of Economics},
  volume={129},
  number={1},
  pages={435--476},
  year={2013}
}

@article{
    johnson2011desegregation, 
    title="{Long-run Impacts of School Desegregation \& School Quality on Adult Attainments}",
    author={Johnson, Rucker C.},
    journal={NBER Working Paper No. 16664}, 
    year={2011}
}

@article{wells1994integration,
  title="{Perpetuation Theory and the Long-Term Effects of School Desegregation}",
  author={Wells, Amy Stuart and Crain, Robert L.},
  journal={Review of Educational Research},
  volume={64},
  number={4},
  year={1994}
}

@article{davies2011intergroup,
  title="{Cross-Group Friendships and Intergroup Attitudes: A Meta-Analytic Review}",
  author={Davies, K. and Tropp, L. R. and Aron, A. and Pettigrew, T. F. and Wright, S. C.},
  journal={Personality and Social Psychology Review},
  volume={15},
  number={4},
  pages={332--351},
  year={2011}
}

@misc{wells2016benefit,
  author = {Wells, Amy Stuart and Fox, L. and Cordova-Cobo, D.},
  title = {How Racially Diverse Schools and Classrooms Can Benefit All Students},
  year = {2016},
  howpublished = {The Century Foundation},
  url = {https://tcf.org/content/report/how-racially-diverse-schools-and-classrooms-can-benefit-all-students/},
  urldate = {2022-02-07}
}

@article{monarrez2021boundaries,
  title="{School Attendance Boundaries and the Segregation of Schools in the US}",
  author={Monarrez, Tomas},
  journal={Working paper},
  year={2021}
}

@article{richards2014gerrymandering,
  title="{The Gerrymandering of School Attendance Zones and the Segregation of Public Schools: A Geospatial Analysis}",
  author={Richards, Meredith P.},
  journal={American Educational Research Journal},
  volume={51},
  number={6},
  year={2014}
}

@article{saporito2016shapes,
  title="{Do Irregularly Shaped School Attendance Zones Contribute to Racial Segregation or Integration?}",
  author={Saporito, S. and Riper, D. V.},
  journal={Social Currents},
  volume={3},
  number={1},
  year={2016}
}

@misc{kahlenberg2016integration,
  author = {Kahlenberg, Richard D.},
  title = {School Integration in Practice: Lessons from Nine Districts},
  year = {2016},
  howpublished = {The Century Foundation},
  url = {https://tcf.org/content/report/school-integration-practice-lessons-nine-districts/},
  urldate = {2022-02-07}
}

@article{monarrez2022charters,
  title="{The Effect of Charter Schools on School Segregation}",
  author={Monarrez, Tomas and Kisida, Brian and Chingos, Matthew},
  journal={American Economic Journal: Economic Policy},
  volume={14},
  number={1},
  pages={1--42},
  year={2022}
}

@article{campos2022zones,
  title="{The Impact of Neighborhood School Choice: Evidence from Los Angeles' Zones of Choice}",
  author={Campos, Christopher and Kearns, Caitlin},
  journal={SSRN Working Paper},
  year={2022}
}

@misc{monarrez2021urban,
  author = {Monarrez, Tomas and Chien, Carina},
  title = {Dividing Lines: Racially Unequal School Boundaries in US Public School Systems},
  year = {2021},
  howpublished = {The Urban Institute},
  url = {https://www.urban.org/sites/default/files/publication/104736/dividing-lines-racially-unequal-school-boundaries-in-us-public-school-systems.pdf},
  urldate = {2022-02-07}
}

@article{fiel2013boundaries,
  title="{Decomposing School Resegregation: Social Closure, Racial Imbalance, and Racial Isolation}",
  author={Fiel, Jeremy E.},
  journal={American Sociological Review},
  volume={78},
  number={5},
  year={2013}
}

@inproceedings{
  mota2021districts,
  title={Fair Partitioning of Public Resources: Redrawing District Boundary to Minimize Spatial Inequality in School Funding},
  author={Mota, Nuno and Mohammadi, Negar and Dey, Palash and Gummadi, Krishna P. and Chakraborty, Abhijnan},
  booktitle={WWW '21: Proceedings of the Web Conference},
  year={2021}
}

@article{caro2004integer,
  title="{School redistricting: embedding GIS tools with integer programming}",
  author={Caro, F. and Shirabe, T. and Guignard, M. and Weintraub, A.},
  journal={Journal of the Operational Research Society},
  volume={55},
  pages={836--849},
  year={2004}
}

@misc{congressional2021contiguity,
  title = {Congressional Redistricting Criteria and Considerations},
  year = {2021},
  howpublished = {Congressional Research Service},
  url = {https://crsreports.congress.gov/product/pdf/IN/IN11618},
  urldate = {2022-02-07}
}

@article{becker2020redistricting,
  title = "{Redistricting Algorithms}",
  author = {Becker, Amariah and Solomon, Justin},
  journal = {arXiv: 2011.09504},
  year = {2020}
}

@book{
    pascal1989cp, 
    author = {Van Hentenryck, P.},
    title = "{Constraint Satisfaction in Logic Programming}",
    year={1989},
    publisher={MIT Press}
}

@misc{cpsat,
  author = {Google OR-Tools},
  title = {CP-SAT Solver},
  year = {2022},
  publisher = {Google},
  url = {https://developers.google.com/optimization/cp/cp_solver},
}

@misc{mcmillan2018boundaries,
  author = {McMillan, Susan M.},
  title = {Common Practices in Changing School Attendance Zone Boundaries},
  year = {2018},
  howpublished = {Educational Data Systems},
  url = {https://eddata.com/wp-content/uploads/2018/10/Common-Practices-in-Changing-School-Attendance-Zone-Boundaries.pdf},
  urldate = {2022-02-08}
}

@misc{
    whitehurt2017segregation, 
    author = {Whitehurt, Grover J.},
    title = "{New evidence on school choice and racially segregated schools}",
    journal = {Brookings},
    url = {https://www.brookings.edu/research/new-evidence-on-school-choice-and-racially-segregated-schools/},
    year={2017},
    urldate = {2022-02-08}
}

@misc{totenberg2007supreme,
  author = {Totenberg, Nina},
  title = {Supreme Court Quashes School Desegregation},
  year = {2007},
  howpublished = {NPR},
  url = {https://www.npr.org/templates/story/story.php?storyId=11598422},
  urldate = {2022-02-08}
}

@misc{mcps2021analysis,
  title = {Montgomery County Public Schools Districtwide Boundary Analysis},
  year = {2021},
  howpublished = {Montgomery County Public Schools},
  url = {https://www.montgomeryschoolsmd.org/uploadedFiles/departments/publicinfo/Boundary_Analysis/BoundaryAnalysis_Final%20Report.pdf},
  urldate = {2022-02-08}
}

@book{
    tatum1997cafeteria, 
    author = {Tatum, Beverly D.},
    title = "{Why are all the Black kids sitting together in the cafeteria? And other conversations about race.}",
    year={1997},
    publisher={Basic Books: New York}
}

@article{moody2001tracking,
  title="{Race, School Integration, and Friendship Segregation in America}",
  author={Moody, James},
  journal={American Journal of Sociology},
  volume={107},
  number={3},
  pages={679--716},
  year={2001}
}

@article{
    kane2005housing, 
    title="{School Quality, Neighborhoods and Housing Prices: The Impacts of school Desegregation}",
    author={Kane, Thomas J. and Staiger, Douglas O. and Riegg, Stephanie K.},
    journal={NBER Working Paper No. 11347}, 
    year={2005}
}

@misc{bridges2016eden,
  author = {Bridges, Kim},
  title = {Eden Prairie Public Schools: Adapting to Demographic Change in the Suburbs},
  year = {2016},
  howpublished = {The Century Foundation},
  url = {https://tcf.org/content/report/eden-prairie-public-schools/},
  urldate = {2022-02-08}
}

@article{black1999housing,
  title="{Do Better Schools Matter? Parental Valuation of Elementary Education}",
  author={Black, Sandra E.},
  journal={The Quarterly Journal of Economics},
  volume={114},
  number={2},
  pages={577--599},
  year={1999}
}

@misc{baltimore2019,
  author = {Baltimore Sun Staff},
  title = {In Howard County, a ‘courageous’ plan to redraw school boundaries tests community’s commitment to diversity},
  year = {2019},
  howpublished = {The Baltimore Sun},
  url = {https://www.baltimoresun.com/education/bs-md-howard-school-redistricting-20190906-xhzkmkf2zvgcxdkbd3vqdanblm-story.html},
  urldate = {2022-02-08}
}

@article{massey1988segregation, 
    title="{The Dimensions of Residential Segregation}",
    author={Massey, D. S. and Denton, N. A.},
    journal="{Social Forces}", 
    volume={67},
    number={2},
    pages={281--315},
    year={1988}
}

@incollection{friedman2013vsd,
  author      = "Friedman, Batya and Kahn Jr., Peter H. and Borning, Alan",
  title       = "Value Sensitive Design and Information Systems",
  editor      = "Doorn N. and Schuurbiers D. and van de Poel I. and Gorman M.",
  booktitle   = "Early engagement and new technologies: Opening up the laboratory",
  publisher   = "Springer",
  address     = "Dordrecht",
  year        = 2013,
  pages       = "55--95",
  chapter     = 4,
}

@article{
    kensing1998pd, 
    title="{Participatory Design: Issues and Concerns}",
    author={Kensing, Finn and Blomberg, Jeanette},
    journal={Computer Supported Cooperative Work (CSCW)}, 
    volume={7},
    pages={167--185},
    year={1998}
}

@article{
    frankenberg2011polls, 
    title="{The Polls--Trends:  School Integration Polls}",
    author={Frankenberg, Erica and Jacobsen, Rebecca},
    journal={Public Opinion Quarterly}, 
    volume={75},
    number={4},
    pages={788--811},
    year={2011}
}

@book{
    delmont2016busing, 
    author = {Delmont, Matthew F.},
    title = "{Why Busing Failed: Race, Media, and the National Resistance to School Desegregation}",
    year={2016},
    publisher={University of California Press}
}

@article{
    reber2005flight, 
    title="{Court-Ordered Desegregation: Successes and Failures Integrating American Schools since Brown versus Board of Education}",
    author={Reber, Sarah J.},
    journal={Journal of Human Resources}, 
    volume={40},
    number={3},
    pages={559--590},
    year={2005}
}

@misc{geverdt2018sabs,
  author = {Geverdt, Doug},
  title = {School Attendance Boundary Survey (SABS) File Documentation: 2015-16 (NCES 2018-099)},
  year = {2018},
  howpublished = {U.S. Department of Education},
  url = {http://nces.ed.gov/pubsearch},
  urldate = {2022-02-09}
}

@online{cpsat2020youtube,
    title = {CPAIOR 2020 Master Class: Constraint Programming},
    date = {2020},
    organization = {Youtube},
    author = {Perron, Laurent and Didier, Frederic},
    url = {https://youtu.be/lmy1ddn4cyw},
}

@Manual{r2021boot,
    title = {boot: Bootstrap R (S-Plus) Functions},
    author = {Canty, Angelo and Ripley, B. D. },
    year = {2021},
    note = {R package version 1.3-28},
}

@Book{davison1997boot,
    title = {Bootstrap Methods and Their Applications},
    author = {Davison, A. C. and Hinkley, D. V.},
    publisher = {Cambridge University Press},
    address = {Cambridge},
    year = {1997},
    note = {ISBN 0-521-57391-2},
    url = {http://statwww.epfl.ch/davison/BMA/},
}

@article{
    abdulkadiroglu2019parents, 
    title="{Do Parents Value School Effectiveness?}",
    author={Abdulkadiroglu, A. and Pathak, P. A. and Schellenberg, J. and Walters, C. R.},
    journal={NBER Working Paper No. 23912}, 
    year={2019}
}

@article{
    pathak2021demand, 
    title="{How well do structural demand models work? Counterfactual predictions in school choice}",
    author={Pathak, Parag A. and Shi, Peng},
    journal={Journal of Econometrics}, 
    volume={222},
    number={1},
    pages={161--195},
    year={2021}
}

@article{
    gilraine2018class, 
    title="{Education Reform in General Equilibrium: Evidence from California's Class Size Reduction}",
    author={Gilraine, M. and Macartney, H. and McMillan, R.},
    journal={NBER Working Paper No. 24191}, 
    year={2018}
}

@article{
    zhang2008flight, 
    title="{White Flight in the Context of Education: Evidence from South Carolina}",
    author={Zhang, Haifeng},
    journal={Journal of Geography}, 
    volume={107},
    number={6},
    pages={236--245},
    year={2008}
}

@incollection{
    torres2020integration,
    title = {Do Parents Really Want School Integration?},
    author = {Torres, E. and Weissbourd, R.},
    year = {2020},
    publisher = {Harvard Graduate School of Education},
    url = {https://static1.squarespace.com/static/5b7c56e255b02c683659fe43/t/5e30a6280107be3cf98d15e6/1580246577656/Do+Parents+Really+Want+School+Integration+2020+FINAL.pdf}
}

@article{
    card2008tipping, 
    title="{Tipping and the Dynamics of Segregation}",
    author={Card, D. and Mas, A. and Rothstein, J.},
    journal="{The Quarterly Journal of Economics}", 
    volume={123},
    number={1},
    pages={177--218},
    year={2008}
}

@article{
    schelling1971segregation, 
    title="{Dynamic Models of Segregation}",
    author={Schelling, T. C.},
    journal={Journal of Mathematical Sociology}, 
    volume={1},
    pages={143--186},
    year={1971}
}

@article{
    billingham2016parents, 
    title="{School Racial Composition and Parental Choice: New Evidence on the Preferences of White Parents in the United States}",
    author={Billingham, Chase M. and Hunt, Matthew O.},
    journal={Sociology of Education}, 
    volume={89},
    number={2},
    year={2016}
}

@article{
    hailey2021parents, 
    title="{Racial Preferences for Schools: Evidence from an Experiment with White, Black, Latinx, and Asian Parents and Students}",
    author={Hailey, Chantal A.},
    journal={Sociology of Education}, 
    year={2021}
}

@article{
    hall2017migration, 
    title="{Latino Students and White Migration from School Districts, 1980-2010}",
    author={Hall, Matthew and Hibel, Jacob},
    journal={Social Problems}, 
    volume={64},
    number={4},
    pages={457--475},
    year={2017}
}

@article{
    charles2003seg, 
    title="{The Dynamics of Racial Residential Segregation}",
    author={Charles, Camille Zubrinsky},
    journal={Annual Review of Sociology}, 
    volume={29},
    pages={167--207},
    year={2003}
}

@article{
    iceland2010households, 
    title="{Racial and Ethnic Residential Segregation and Household Structure: A Research Note}",
    author={Iceland, John and Goyette, Kimberly A. and Nelson, Kyle Ann and Chan, Chaowen},
    journal={Social Science Research}, 
    volume={39},
    number={1},
    pages={39--47},
    year={2010}
}

@article{
    pager2005bias, 
    title="{Walking the Talk? What Employers Say Versus What They Do}",
    author={Pager, Devah and Quillian, Lincoln},
    journal={American Sociological Review}, 
    volume={70},
    number={3},
    year={2005}
}

@article{
    candipan2019neighborhoods, 
    title="{Neighbourhood change and the neighbourhood-school gap}",
    author={Candipan, Jennifer},
    journal={Urban Studies}, 
    volume={56},
    number={15},
    year={2019}
}

@article{
    rich2021seg, 
    title="{Segregated Neighborhoods, Segregated Schools: Do Charters Break a Stubborn Link? }",
    author={Rich, Peter and Candipan, Jennifer and Owens, Ann},
    journal={Demography}, 
    volume={58},
    number={2},
    pages={471--498},
    year={2021}
}

@article{harwell2010frl,
  title="{Student Eligibility for a Free Lunch as an SES Measure in Education Research}",
  author={Harwell, Michael and LeBeau, Brandon},
  journal={Educational Researcher},
  volume={39},
  number={2},
  year={2010}
}

@article{jakubs1977dissim,
  title="{Residential Segregation: The Taueber Index Reconsidered}",
  author={Jakubs, John F.},
  journal={Journal of Regional Science},
  volume={17},
  number={2},
  year={1977},
  pages={281--283}
}

@article{james1985seg,
  title="{Measures of Segregation}",
  author={James, David R. and Taeuber, Karl E.},
  journal={Sociological Methodology},
  volume={15},
  year={1985},
  pages={1--32}
}

@article{mehrotra1998contiguity,
  title="{An Optimization Based Heuristic for Political Districting}",
  author={Mehrotra, Anuj and Johnson, Ellis L. and Nemhauser, George L.},
  journal={Management Science},
  volume={44},
  number={8},
  pages={1021--1166},
  year={1998}
}

@misc{chang2018shsat,
  author = {Chang, Alvin},
  title = {The fraught racial politics of entrance exams for elite high schools},
  year = {2018},
  howpublished = {Vox},
  url = {https://www.vox.com/2018/6/14/17458710/new-york-shsat-test-asian-protest},
  urldate = {2022-04-21}
}

@article{
    liggett1973optimization, 
    title="{The Application of an Implicit Enumeration Algorithm to the School Desegregation Problem}",
    author={Liggett, Robin Segerblom},
    journal={Management Science}, 
    volume={20},
    number={2},
    pages={159--168},
    year={1973}
}

@article{
    clarke1968optimization, 
    title="{An operations research approach to racial desegregation of school systems}",
    author={Clark, S. and Surkis, J.},
    journal={Socio-Economic Planning Sciences}, 
    volume={1},
    number={3},
    pages={259--272},
    year={1968}
}

@inproceedings {allman2021sfusd,
author={Allman, Maxwell and Ashlagi, Itai and Lo, Irene and Mentzer, Katherine},
booktitle={Stanford Institute for Computational \& Mathematical Engineering Xpo Research Symposium},
title={Poster: Designing School Choice for Diversity in the San Francisco Unified School District},
year={2021}
}

@article{
    chetty2022socialcapitalII, 
    author = {Chetty, Raj and Jackson, Matthew O. and Kuchler, Theresa and Stroebel, Johannes et al.},
    title = {Social capital II: determinants of economic connectedness},
    year={2022},
    journal={Nature},
    volume={608},
    pages={122--134}
}

@misc{owens2022normalized,
  author={Owens, Ann and reardon, sean f. and Kalogrides, Demetra and Jang, Heewon and Tom, Thalia},
  title = {Measuring Segregation with the Normalized Exposure Index},
  year = {2022},
  howpublished = {Research Brief},
  url = {https://socialinnovation.usc.edu/wp-content/uploads/2022/05/Measuring-Segregation-with-the-Normalized-Exposure-Index_Rnd6.pdf},
  urldate = {2022-09-01}
}

@article{
    winship1978dissim, 
    author = {Winship, Christopher},
    title = {The Desirability of Using an Index of Dissimilarity or any Adjustment of it for Measuring Segregation: Reply to Falk, Cortese, and Cohen},
    year={1978},
    journal={Social Forces},
    volume={57},
    number={2},
    pages={717--720}
}

@article{
    reardon2002multigroup, 
    author = {Reardon, Sean F. and Firebaugh, Glenn},
    title = {Measures of Multigroup Segregation},
    year={2002},
    journal={Sociological Methodology},
    volume={32},
    pages={33--67}
}

@misc{houlgrave2021choice,
  author = {Houlgrave, Bryon},
  title = {COVID-19 May Energize Push for School Choice in States. Where That Leads Is Unclear},
  year = {2021},
  howpublished = {EdWeek},
  url = {https://www.edweek.org/policy-politics/covid-19-may-energize-push-for-school-choice-in-states-where-that-leads-is-unclear/2021/01},
  urldate = {2022-09-02}
}

@misc{nces2021choicefacts,
  author={US Department of Education},
  title = {Fast Facts: Public school choice programs},
  year = {2021},
  howpublished = {Digest of Education Statistics, 2019 (NCES 2021-009)},
  url = {https://nces.ed.gov/fastfacts/display.asp\?id=6},
  urldate = {2022-09-02}
}

@article{
    schachner2022LA, 
    author = {Schachner, Jared N.},
    title = {Racial Stratification and School Segregation in the Suburbs: Evidence from Los Angeles County},
    year={2022},
    journal={Social Forces},
    volume={101},
    number={1},
    pages={309--340}
}

@article{
    bischoff2020imbalance, 
    author = {Bischoff, Kendra and Tach, Laura},
    title = {School Choice, Neighborhood Change, and Racial Imbalance Between Public Elementary Schools and Surrounding Neighborhoods},
    year={2020},
    journal={Sociological Science}
}

@article{
    hanushek2004switching, 
    author = {Hanushek, Eric A. and Kain, John F. and Rivkin, Steven G.},
    title = {Disruption versus Tiebout improvement: the costs and benefits of switching schools},
    year={2004},
    journal={Journal of Public Economics},
    volume={88},
    pages={1721--1746}
}

@article{
    schwartz2017switching, 
    author = {Schwartz, Amy E. and Stiefel, Leanna and Cordes, Sarah A.},
    title = {Moving Matters: The Causal Effect of Moving Schools on Student Performance},
    year={2017},
    journal={Education Finance and Policy},
    volume={12},
    number={4},
    pages={419--446}
}

@article{
    nielsen2020denmark, 
    title="{Attendance boundary policies and the limits to combating school segregation}",
    author={Bjerre-Nielsen, Andreas and Gandil, Mikkel H.},
    journal={Working paper}, 
    year={2020}
}




\end{document}